# ACCURACY OF GARMIN GPS RUNNING WATCHES OVER REPETITIVE TRIALS ON THE SAME ROUTE

Joe Dumas

Department of Computer Science and Engineering,
University of Tennessee at Chattanooga, Chattanooga, Tennessee, USA

## ABSTRACT

*Many runners use watches incorporating Global Positioning System technology to track their workouts. These devices can be valuable training aids, but they have limitations. For several reasons including variations in satellite position, environmental factors, and design decisions made by the manufacturer, GPS-enabled watches can produce position measurement errors. These can result in incorrect estimations of total distance covered as well as running pace. This study examined the accuracy of three Garmin running watches of different technological generations using repetitive trials, over several years, by the same runner over the same route. The older watches, a Forerunner 205 and a Forerunner 220, showed similar accuracy when traversing the route. The newer generation watch, a Forerunner 45S, was found to be significantly less accurate in terms of both the trueness and precision of its distance measurements. This may indicate that Garmin, in competition with other manufacturers of similar devices, has chosen in recent years to prioritize device miniaturization and battery life over accuracy.*

## 1. INTRODUCTION

Many runners from the recreational level up to elite professional athletes use watches incorporating Global Positioning System (GPS) technology to track their workouts [1]. While these devices are capable of collecting and displaying a variety of data, their most fundamental function is to record the geographic location of the wearer at discrete intervals of time. From the time/position data, the watch can compute relevant parameters such as total time elapsed for the workout, distance traveled, pace (per mile or kilometer), etc. At the conclusion of a workout, the user has the opportunity to save the workout data to a file stored on the device; this file can later be transferred to a computer for further analysis and archival.

GPS technology uses a collection of satellites orbiting the earth to triangulate the user's position based on the length of time it takes for signals from each satellite to reach the receiver (in this case, the sports watch) [2]. GPS was invented in the 1970s, with a full constellation of satellites in orbit by 1993, but originally was restricted to use by the United States military [3]. Even after civilian use was allowed, intentional degradation of the signal available to non-military devices (known as Selective Availability) limited the system's usefulness [4]. It was only after Selective Availability was discontinued in 2000 that civilian-owned devices could utilize GPS with full accuracy. However, even unrestricted GPS is not a perfectly accurate position tracking system; there are certain types of errors that are inherent to the system, and in addition, specific GPS-enabled devices can contribute their own sources of error [5, 6]. These various types of GPS errors can contribute to inaccurate measurement of the user's position, which can then translate (in the case of a running watch) to inaccurate calculations of the athlete's pace and distance covered. This last type of error – misestimation of distance traveled – is the subject of this study.

Several companies currently manufacture GPS-enabled watches for use by runners and other athletes. One of the most popular GPS watchmakers is Garmin Ltd., headquartered in Olathe, Kansas, USA [5]. Garmin has been manufacturing various types of GPS receivers (marine, automotive, avionics) since 1990 and entered the fitness market in 2003 with the introduction of the first generation of its Forerunner series of watches [7]. While the early models were large and "clunky" and had only minimal features and short battery life, Garmin's fitness watches have improved markedly over the years – becoming smaller, more feature-rich, and gradually increasing the amount of time they can operate before they require a charge.

It is relatively easy to find – both in the context of product reviews as well as more formal studies – comparisons of devices that are (or were) competitors in the fitness market at a given time. Many such reviews focus more on product appearance, features, ease of use, etc. than on the accuracy of the data produced by the devices. However, there are few if any studies that consider the evolution of a given manufacturer's products over an extended period of time, particularly with respect to whether or not advances in technology result in improved accuracy. The main contribution of this paper is to examine that question in the specific context of Garmin's GPS running watches.

The rest of this paper will provide a brief background on the devices studied and the author's experience in using them, the types of errors that are commonly experienced when using GPS-enabled devices, and the types of design tradeoffs that may be made by device manufacturers who are trying to maximize sales in a competitive global market. This will be followed by a description of the methodology used to collect and analyze data to assess the accuracy of the devices studied, with respect to running activity distance measurements. The collected data will be summarized and hypotheses regarding the relative and absolute accuracy of the devices will be presented. Statistical analysis techniques will be used to support or refute the hypotheses. The results will be discussed and relevant conclusions will be presented.

## 2. BACKGROUND

The author is a university professor with a Ph.D. in Computer Engineering and has been an avid runner for many years. Since 2010 he has owned a series of Forerunner GPS running watches made by Garmin. He has used these watches to track races of various distances as well as training runs and has saved all data files from every GPS-tracked run. Over the years, the author has developed the habit of running a variety of convenient routes that start and end at his home. Some of those routes have become "favorites" that the author runs again and again, giving him a large collection of logged data from certain routes. The author has also noted that no matter how consistently he runs the same route, there is some variation in the distance reported by the GPS watch at the completion of the run. This led him to the idea of choosing a frequently traveled route, collecting the data files from every time he has ever run that route, and analyzing the data (specifically the total recorded workout distance) to try to determine whether some watch models are more accurate than others.

The three specific GPS watch models owned and used by the author are the Garmin Forerunner 205 (introduced by Garmin in January 2006, used December 2010 – October 2015), the Forerunner 220 (introduced September 2013, used October 2015 – December 2019), and the Forerunner 45S (introduced April 2019, used December 2019 - present) [7]. These models represent the second, fourth, and sixth generations of Garmin fitness devices. Their characteristics are summarized in Table 1.

Table 1. Comparison of GPS Running Watches Studied.

| Model | Forerunner 205 | Forerunner 220 | Forerunner 45S |
|---|---|---|---|
| Year Introduced | 2006 | 2013 | 2019 |
| Dimensions (mm) | 53.3 x 68.6 x 17.8 | 45.0 x 45.0 x 12.5 | 39.5 x 39.5 x 11.4 |
| Weight (g) | 77.0 | 40.7 | 32.0 |
| Rated Battery Life (h) | 10 | 10 | 13 |
| GPS Chipset | SIRFstarIII | MediaTek MT3333 | Sony CXD5603GF |

There are several factors that can cause distances measured by GPS devices to be incorrect. Most basically, GPS devices do not measure the wearer's position continuously, but only at discrete intervals of time [6]. For running watches, the typical sampling interval between position measurements is around one second [8], but some models allow the interval to be increased to several seconds, or even up to one minute, in order to extend battery life for long-duration activities. Software running on the watch then calculates the distance between each successive pair of points and accumulates the total distance. Thus, the run is treated as a collection of straight-line segments rather than a continuous activity. However, no runner will run in a perfectly straight line, so estimating the distance in this way will invariably result in some amount of error. Generally, the greater the time interval between sampling points, the more error is likely to occur.

Additionally, position location using GPS is subject to errors due to timing limitations, minor variations in satellite position, atmospheric conditions, nearby objects (*e.g.*, hills, tall buildings), etc. [6]. Consumer-grade GPS devices are generally only considered to be accurate to approximately 3 meters (10 feet) [5], depending on the number of satellites acquired by the device (more satellites generally result in a more accurate "fix" of position) [2]. The general trend observed by runners and seen in the literature is that, for most routes, the distance indicated by a GPS watch is slightly greater than the actual distance traveled (Lovett [9] reports 1% overestimation of distance, and therefore pace, as typical). However, in pathological cases (*e.g.*, hilly trail runs with many tight switchbacks) GPS interpolation may "cut off" some of the curves and replace them with straight line segments, resulting in displaying a distance shorter than the true distance traveled.

The inherent inaccuracies described above can be exacerbated by design tradeoffs made by device manufacturers. Manufacturers such as Garmin, Polar, Suunto, and TomTom have increasingly prioritized minimizing device size and weight while maximizing battery life as selling points for their devices [9], but these design goals may be at odds with one another (*e.g.*, physically smaller batteries typically have lower milliampere hour ratings), forcing compromises to be made. Maximizing battery life is particularly important in marketing GPS watches to marathon and ultramarathon runners whose races and training runs may last several hours or more. Since GPS location measurements and the associated computations place greater demands on the device, requiring more current from the battery, a common way to extend run time on a single charge is to increase the GPS sampling interval. This represents a tradeoff accepting decreased accuracy in return for longer duration of operation of a device of a given size.

## 3. METHODOLOGY

The author chose a route that he had run repeatedly and consistently over time and reviewed over ten years' worth of run data within the Garmin Training Center program to identify all runs that had used that route. He then gathered the data files from each of those runs for analysis. The data logged by the Forerunner 205 were stored in TCX (Training Center XML) files which is a human-readable format, so it was easy to locate the total distance traveled (recorded in meters) at the end of each file. The data logged by the newer watches (Forerunner 220 and 45S) were stored in FIT (Flexible and Interoperable data Transfer) files, a Garmin proprietary binary format that is not human-readable. The author located an online format conversion tool [10] and used it to convert each FIT file to a TCX file, such that the recorded distance of each workout could be determined. The date and reported distance of each run were then entered into an Excel spreadsheet for statistical analysis.

To evaluate the accuracy of the GPS watches, it was necessary to establish a true, or at least as nearly accurate as possible, distance for the chosen route. A detailed survey by a professional land surveyor would likely give the most accurate distance, but that approach was not financially feasible for this project. The next best approach would have been to hire a USA Track & Field (USATF) road course certifier to measure the route with a calibrated bicycle wheel and a Jones counter (this is the way race courses are measured) [9]; however, this also proved impractical within the scope of the project. Instead, the author used three different online tools, each based on Geographical Information System (GIS) data, to estimate the actual length of the route. He traced the route on each platform based on known starting and ending points and the streets traversed. OnTheGoMap.com gave a distance of 5.68 km for the route [11]. MapMyRun.com showed the route length to be 5.69 km [12]. Gmap-pedometer.com, which was recommended by Lovett [9] as "accurate to within 0.1 percent," measured the route to be 5.685 km [13]. The route as mapped using this site is shown in Figure 1. This last distance value (5685 meters), which also represented the mean and median of the three GIS-based distance estimates, was taken to be the nominal distance covered while running the route. It was added to the Excel spreadsheet used to analyze the collected data.

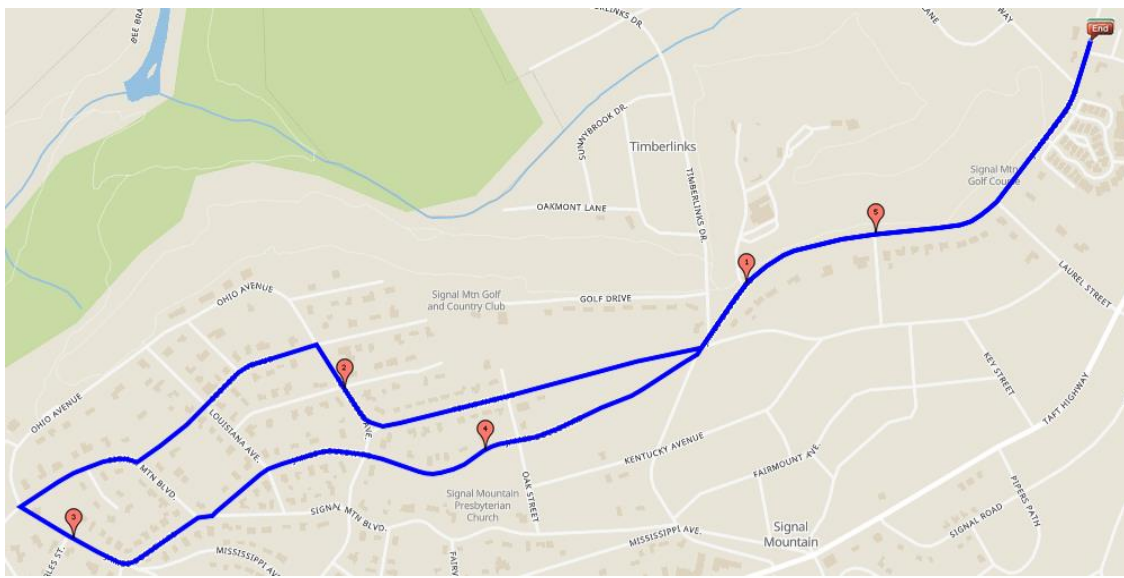

Figure 1.  Route mapped using GIS mapping tool

Statistical analysis of the data from each GPS device focused not only on the mean distance reported by each device, but also on the variability of the distance measurements. Savage [14] evaluated GPS running watches for accuracy using "the ISO 5725 definition of accuracy as the combination of trueness and precision." He noted that "we can look at trueness by measuring the average … length and precision by measuring the standard deviation." In other words, trueness is indicated by how close the average (mean) of the distance measurements reported by a device is to the actual distance covered. A small average difference from the actual distance indicates high trueness, while a small standard deviation of the distance values indicates high precision. An ideal device would exhibit high accuracy by excelling in both trueness and precision. A poor device would be inferior in both these aspects of accuracy. It is also possible for a device of intermediate quality to have high trueness but low precision, or *vice versa*. Which of those scenarios would be more problematic would depend on a given user's priorities and use cases. The analysis to follow compares the selected devices considering both aspects of accuracy.

## 4. RESULTS

The number of data points collected varied considerably among the three watches due to the length of time each one was used vs. the author's frequency of running the chosen route. (He ran it less often in the earlier years and more frequently later on.) A total of 18 data points were recorded by the Forerunner 205, while the 220 and 45S recorded 72 and 59 runs, respectively. The data collected from each watch are summarized in Table 2 below.

Table 2. Summary of Data Collected from GPS Watches.

| Model | Forerunner 205 | Forerunner 220 | Forerunner 45S |
|---|---|---|---|
| N (# observations) | 18 | 72 | 59 |
| Mean distance (m) | 5668.495 | 5666.798 | 5740.030 |
| Std. dev. (m) | 17.027 | 16.624 | 29.627 |
| Minimum value (m) | 5644.769 | 5629.406 | 5668.852 |
| Maximum value (m) | 5713.828 | 5716.930 | 5804.433 |

The first data analysis focused on comparing the watches vs. each other. The null hypothesis of "no significant difference between distance reported by different watches" was tested against the alternative hypotheses that the newer watches would be less accurate, due to the manufacturer's emphasis, in recent years, on maximizing battery life rather than accuracy. Since battery life is typically maximized by increasing the time interval between GPS position readings and since (as previously noted) doing so tends to make the device "measure long," the specific hypotheses being tested were that each newer watch model would report a greater distance than previous model(s). Specifically:

$H_1$: Forerunner 220 distance > Forerunner 205 distance
$H_2$: Forerunner 45S distance > Forerunner 205 distance
$H_3$: Forerunner 45S distance > Forerunner 220 distance

Testing of the hypotheses was done using a built-in Excel data analysis function that performs a two-sample t-test assuming unequal variances. While ideally this test would compare data sets each comprising 30 or more values, the 18 observations obtained from the Forerunner 205 were approximately normally distributed and their variance was not excessive compared to the other data sets. Thus, it is reasonable to apply the t-test to compare the data samples from the three devices. The results of these tests are summarized in Tables 3, 4, and 5 below.

Table 3. Comparison of Forerunner 205 vs. Forerunner 220.

| Forerunner 205 | | | Forerunner 220 | | | t-statistic | *p*-value (one-tailed) |
|---|---|---|---|---|---|---|---|
| N | Mean | Variance | N | Mean | Variance | | |
| 18 | 5668.495 | 289.908 | 72 | 5666.798 | 276.351 | 0.3798947 | 0.3535546 |

Table 4. Comparison of Forerunner 205 vs. Forerunner 45S.

| Forerunner 205 | | | Forerunner 45S | | | t-statistic | *p*-value (one-tailed) |
|---|---|---|---|---|---|---|---|
| N | Mean | Variance | N | Mean | Variance | | |
| 18 | 5668.495 | 289.908 | 59 | 5740.030 | 877.772 | -12.85147 | 9.107E-18 |

Table 5. Comparison of Forerunner 220 vs. Forerunner 45S.

| Forerunner 220 | | | Forerunner 45S | | | t-statistic | *p*-value (one-tailed) |
|---|---|---|---|---|---|---|---|
| N | Mean | Variance | N | Mean | Variance | | |
| 72 | 5666.798 | 276.351 | 59 | 5740.030 | 877.772 | -16.92756 | 1.440E-29 |

It should be pointed out that not all the variability in the collected data is due to the GPS watches themselves. Although the author attempted to run the route as consistently as possible over time, small path deviations (*e.g.*, detours around other pedestrians, parked cars, etc.) were inevitable, and these necessarily resulted in small variations in the actual distance covered on each run. Based on the author's knowledge of the route and his running habits, it is estimated that these deviations were almost always less than 0.01 mile (52.8 feet or approximately 16 m) from the nominal route distance. Any deviations in excess of this amount should be attributed to errors inherent to the GPS location system and/or its implementation in a particular device.

Examining the test results, it is clear that $H_1$ is not supported. The Forerunner 220 did not report greater distances than the Forerunner 205; in fact, its mean distance reading was approximately 1.7 meters *shorter* than the older device. This difference of 0.03% is smaller than the author's estimate of inherent path variability and much too small to be statistically significant.

Conversely, there is strong statistical evidence to support $H_2$ and $H_3$. The mean distance recorded by the Forerunner 45S was more than 70 meters (nearly 1.3%) greater than the distance recorded by either of the other two devices. The tiny *p*-values shown in Tables 4 and 5 indicate an infinitesimal likelihood (much, much less than 1%) that the large differences between the distances reported by the Forerunner 45S vs. the other two devices could have happened by chance. It is also worth noting that while the data from the two older watches showed similar variability (only about 0.4 m difference between the first two standard deviations reported in Table 2), the data obtained from the Forerunner 45S were much more variable (with a standard deviation nearly twice as great as the other two devices).

The second data analysis focused on comparing the distance reported by the three watches vs. the nominal route distance obtained from the GIS plotting sites. The null hypothesis in this case is "no significant difference between distance reported by GPS watches and distance measured using GIS." The author's initially intended alternative hypotheses were that each of the watches would report a distance *greater* than the GIS measured distance. This prediction arose from the fact that GPS watches "measuring long" is a trend noted in several places in the literature [6, 9, 15]. In particular, Johansson *et al* [15] found Garmin Forerunner models (generically) to overestimate distance by an average of 0.6% during a 56 km ultramarathon race. However, given the GIS-based nominal distance for the route of 5685 meters, it was immediately clear that both the Forerunner 205 and Forerunner 220 averaged reporting *shorter* distances (by approximately 16.5 and 18.2 meters, respectively) than the estimated actual distance covered. Only the Forerunner 45S recorded a mean distance reading greater (by about 55 meters) than the GIS-measured distance for the route.

Because of the above observation, the alternative hypotheses were adjusted to reflect the possibility that a watch could potentially report a distance either significantly *greater or less* than the distance established using GIS mapping. Therefore, the specific alternative hypotheses tested were:

$H_4$: Forerunner 205 distance $\neq$ GIS distance
$H_5$: Forerunner 220 distance $\neq$ GIS distance
$H_6$: Forerunner 45S distance $\neq$ GIS distance

In this case the data were analyzed using a one-sample t-test. While Excel does not support this function directly, Howard [16] has devised a way to perform this test using the program's provided functionality for the two-sample t-test assuming unequal variances (used in the previous watch vs. watch comparisons). The tables below present the results of this analysis.

Table 6. Comparison of Forerunner 205 vs. GIS Route Distance.

| Forerunner 205 | | | Distance Estimated via GIS | t-statistic | *p*-value (two-tailed) |
|---|---|---|---|---|---|
| N | Mean | Variance | | | |
| 18 | 5668.495 | 289.908 | 5685 | -4.1126919 | 0.0007265 |

Table 7. Comparison of Forerunner 220 vs. GIS Route Distance.

| Forerunner 220 | | | Distance Estimated via GIS | t-statistic | *p*-value (two-tailed) |
|---|---|---|---|---|---|
| N | Mean | Variance | | | |
| 72 | 5666.798 | 276.351 | 5685 | -9.2906966 | 6.831E-14 |

Table 8. Comparison of Forerunner 45S vs. GIS Route Distance.

| Forerunner 45S | | | Distance Estimated via GIS | t-statistic | *p*-value (two-tailed) |
|---|---|---|---|---|---|
| N | Mean | Variance | | | |
| 59 | 5740.030 | 877.772 | 5685 | 14.2669884 | 1.262E-20 |

From the statistics shown above, there is evidence to support all three of $H_4$, $H_5$, and $H_6$ at considerably greater than a 99% level of confidence. In other words, the data from all three Garmin Forerunner watches differed significantly from the best available estimate of the actual length of the route. The Forerunner 205 had a mean deviation of -0.29% from the GIS distance, while the mean deviation of the Forerunner 220 vs. GIS was -0.32%. Conversely, the Forerunner 45S averaged approximately 0.97% "long" compared to the GIS estimated distance.

## 5. DISCUSSION

From the analysis above, it can be seen that the performance of the older two GPS watches studied (the Forerunner 205 and 220) was essentially equivalent from the standpoint of accuracy. Their mean recorded distances were within 0.03% of each other and their standard deviations of distance were similar. Based on the definition of accuracy as the combination of trueness and precision, these two devices exhibited statistically equal accuracy.

On the other hand, comparing the performance of the Forerunner 45S to that of its predecessors, there is ample statistical evidence to support the assertion that its distance readings are less accurate. Its mean reported distance was nearly 1% greater than the GIS-plotted route and about 1.3% longer than the mean distance indicated by the other two watches – both highly significant observations based on the statistical analysis. In addition, the standard deviation of the distance data produced by the Forerunner 45S was 74% greater than that of the Forerunner 205 and 78% greater vs. the Forerunner 220. Both in terms of trueness and precision, the Forerunner 45S used in this study was less accurate than the older models to which it was compared.

While it is clear that the Forerunner 45S tested was not as accurate, in a relative sense, as its predecessors the 205 and 220, it is more difficult to draw solid conclusions regarding *absolute* accuracy. Based on both the available literature and personal experience with GPS running watches, the author expected all three watches to "measure long," reporting mean distances in excess of the actual distance covered. However, when Geographical Information System (GIS) mapping was used to establish the nominal distance for the route, both the Forerunner 205 and 220 "measured short," with mean recorded distances around 0.3% less than the distance indicated by GIS. The Forerunner 45S, on the other hand, measured roughly 1% long compared to the GIS distance. This is a significant overestimation, and one that is supported by other studies.

While we can be fairly certain that the Forerunner 45S produced distance readings that were greater than "ground truth," further investigation may be needed to establish whether the other devices actually underestimated the distance covered. It may be that errors in the GIS database, or in the way that the mapping sites utilized the GIS data, and/or the author's plotting of the route using the sites, resulted in an incorrect determination of the true route length. The only way to establish this with greater certainty would be to use a more accurate measurement method that would involve physically traversing the route with more precise instrumentation (such as a calibrated wheel). If this could be done, the results depicted in Tables 6, 7, and 8 could be revised using the more accurately established value for the length of the route.

Overall, while some statistically significant results were observed in this study, it is important to realize its limitations. Only three different GPS watch models from a single manufacturer were studied, and it was not feasible to test multiple devices of the same type to evaluate intra-model consistency. Only one running route of a modest distance (just under 5.7 km, or about 3.53 miles) was used in the study; the results may have been different if a longer or shorter route had been used. The chosen route included some hills (resulting in moderate altitude gain/loss, approximately ±83 meters according to OnTheGoMap [11]) and a modest amount of turns and road curvature, any or all of which could have affected the results of the study vs. a flatter (or hillier) route, or one with more straight or curved sections. A more comprehensive study would need to include a wider selection of GPS devices and a variety of running routes of varying lengths and characteristics. It is the author's hope that such studies will follow upon this work.

## 6. CONCLUSIONS

We conclude that the performance of the older two GPS watches studied (the Forerunner 205 and 220) was essentially equivalent from the standpoint of accuracy, in terms of both trueness and precision. Thus, this study provides no support for the notion that between 2006 (when the Forerunner 205 was introduced) and 2013 (the debut of the Forerunner 220), manufacturer Garmin Ltd. sacrificed accuracy in order to make its GPS watches smaller and lighter. Conversely, we conclude that the Forerunner 45S used in this study produced distance readings that were significantly less accurate than its predecessors. This supports the conjecture that, sometime between 2013 and the 2019 introduction of the Forerunner 45S model, Garmin may have chosen to de-emphasize accuracy in order to continue to miniaturize its running watches and increase their battery life.

## AUTHOR

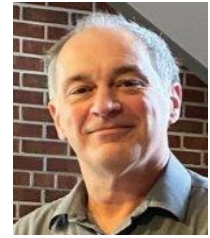

**Joe Dumas** earned the Ph.D. in Computer Engineering from the University of Central Florida in 1993. Dr. Dumas is a UC Foundation Professor in the Computer Science and Engineering Department at the University of Tennessee at Chattanooga. He was chosen Outstanding Computer Science Teacher in 1998, 2002, and 2009. Dr. Dumas' areas of interest include computer architecture, embedded systems, virtual reality, and real-time, human-in-the-loop simulation. He is an avid runner and has completed 70 races of marathon or longer distance. In addition to running, he enjoys tennis and downhill skiing. Joe Dumas resides in Signal Mountain, Tennessee, USA with his wife of nearly 30 years, Chereé.